\documentclass[universe,article,accept,pdftex,moreauthors]{Definitions/mdpi} 

\firstpage{1} 
\pubvolume{1}
\issuenum{1}
\articlenumber{0}
\pubyear{2025}
\copyrightyear{2025}
\externaleditor{Firstname Lastname}
\datereceived{ } 
\daterevised{ } % Comment out if no revised date
\dateaccepted{ } 
\datepublished{ } 
\hreflink{https://doi.org/} % If needed use \linebreak
\newcommand{\sss[1]}{_{_{#1}}}

\Title{A Scenario for Origin of Global 4 mHz Oscillations in Solar Corona}

\TitleCitation{A Scenario for Origin of Global 4 mHz Oscillations in Solar Corona}

\Author{Li %MDPI: Please carefully check the accuracy of names and affiliations.
 Xue $^{1,}$*\orcidA{}, Chengliang %MDPI: We revised the author name format, please confirm.
 Jiao $^{2,3,4,}$*\orcidB{} and Lixin Zhang $^{1}$\orcidC{}}

\AuthorNames{Li Xue, Cheng-Liang Jiao and Li-Xin Zhang}

\AuthorCitation{Xue, L.; Jiao, C.; Zhang, L.}
\address{%
$^{1}$ \quad Department of Astronomy, Xiamen Univesity, Xiamen 361005, China; 
\\
$^{2}$ \quad Yunnan Observatories, Chinese Academy of Sciences, 396 Yangfangwang, Guandu District,\linebreak Kunming 650216,  China\\
$^{3}$ \quad Center for Astronomical Mega-Science, Chinese Academy of Sciences 20A Datun Road, Chaoyang District, Beijing 100012, China\\
$^{4}$ \quad Key Laboratory for the Structure and Evolution of Celestial Objects, Chinese Academy of Sciences, 396 Yangfangwang, Guandu District, Kunming 650216, China
}

\corres{Correspondence: lixue@xmu.edu.cn (L.X.); jiaocl@ynao.ac.cn (C.J.)}

\abstract{We establish a spherically symmetric model of solar atmosphere, which consists of the whole chromosphere and low corona below the $1.25$ solar radius. It is a hydrodynamic model with heating in the chromosphere through an artificial energy flux. We performed a series of simulations with our model and found oscillations with a peak frequency of \mbox{$\sim$4 $\rm{mHz}$} in the power spectrum. We confirmed that this resulted from the $p$-mode excited in the transition region and amplified in a resonant cavity situated in the height range $\sim$$4\times10^3$--$2\times10^4$ km. This result is consistent with global observations of Alfv\'enic waves in corona %MDPI: Ref citation is not allowed in Abstract section, please move it to main text.
 and can naturally explain the observational ubiquity of $4\ \rm{mHz}$ without the difficulty of the $p$-mode passing through the acoustic-damping chromosphere. We also confirmed that acoustic shock waves alone cannot heat the corona to the observed temperature, and found mass upflows in the height range $\sim$$7\times10^3$--$7\times10^4$ km in our model, which pumped the dense and cool plasma into the corona and might be the mass supplier for solar prominences.}


\begin{document}

%%%%%%%%%%%%%%%%%%%%%%%%%%%%%%%%%%%%%%%%%%
%\setcounter{section}{-1} %% Remove this when starting to work on the template.
%\section{How to Use this Template}

%The template details the sections that can be used in a manuscript. Note that the order and names of article sections may differ from the requirements of the journal (e.g., the positioning of the Materials and Methods section). Please check the instructions on the authors' page of the journal to verify the correct order and names. For any questions, please contact the editorial office of the journal or support@mdpi.com. For LaTeX-related questions please contact latex@mdpi.com.%\endnote{This is an endnote.} % To use endnotes, please un-comment \printendnotes below (before References). Only journal Laws uses \footnote.

% The order of the section titles is different for some journals. Please refer to the "Instructions for Authors” on the journal homepage.

\section{Introduction}\label{intro}

Since Edl{\'e}n (1943)~\cite{Edlen1943} showed that the coronium line observed in the Sun is actually a highly ionized iron line emitted at temperatures exceeding one million Kelvins, how to heat and sustain the solar corona millions of Kelvins hotter than the surface, i.e. the coronal heating problem, has become one of the fundamental problems in solar physics. Similarly, there is also a chromospheric heating problem and a solar wind heating problem~\cite{Hollweg1985}. Biermann (1946)~\cite{Biermann1946} and Schwarzschild (1948)~\cite{Schwarzschild1948} first proposed that acoustic waves generated by solar granulation could carry energy from the top of the convective zone into the corona. However, this mechanism was first successfully applied to the chromospheric heating problem instead. Ulmschneider~\cite{Ulmschneider1970,Ulmschneider1971a,Ulmschneider1971b} established a time-dependent hydrodynamic model and discussed the acoustic waves generated in the convection zone and the resulting shock waves as the possible fundamental energy transportation and heating mechanism. Stein and Schwartz~\cite{Stein_Schwartz1972,Stein_Schwartz1973} further considered the effects of ionization using Saha's equation~\cite{Saha1920} on the acoustic pulse and periodic wave train. Subsequently, Ulmschneider et al. (1977)~\cite{Ulmschneider_etal_1977}, Kalkofen and Ulmschneider (1977)~\cite{Kalkofen_Ulmscheider_1977}, and Ulmschneider and Kalkofen (1977)~\cite{Ulmschneider_Kalkofen1977} performed systematic research to investigate acoustic waves in the solar atmosphere, and their results supported the short-period acoustic heating theory of the chromosphere. Anderson and Athay~\cite{Anderson1989a,Anderson1989b} improved the empirical loss proposed by Vernazza et al. (1981)~\cite{Vernazza1981} and recomputed the chromospheric cooling fluxes, which were found to be consistent with the observational propagating acoustic wave flux obtained by Endler and Deubner (1983)~\cite{Endler1983,Deubner1988}. Carlsson and Stein (1992)~\cite{Carlsson1992} established the first fully self-consistent one-dimensional radiation--hydrodynamic model to explain the chromospheric bright point phenomenon, and their calculation of acoustic waves reproduced many aspects of the observed Ca II bright point behavior rather well. Fossum and Carlsson (2005)~\cite{Fossum_Carlsson_2005} proposed that the acoustic energy flux is too low to balance the radiative losses in the chromosphere based on space observations with the Transition Region and Coronal Explorer (TRACE)~\cite{TRACE1999,Fossum_Carlsson_2005b}, but their result was later contested by several works~\cite{Wede2007,Cuntz2007,Kalkofen_2007}, who attributed the missing flux to the limited spatial resolution of TRACE. Bello Gonz{\'a}lez et al. (2010)~\cite{Bello2009} gave a detailed discussion on the acoustic wave heating debate, and Bello Gonz{\'a}lez et al. (2010)~\cite{Bello2010} found that the energy flux of acoustic waves in the 5.2--10 mHz range, obtained from the wavelet analysis of high-resolution spectropolarimetric data taken with IMaX/Sunrise and considered to be lower limits, lies within a factor of two of the energy flux needed to balance the radiative losses from the chromosphere according to the estimates of Anderson and Athay (1989)~\cite{Anderson1989b}. Recent observations~\cite{Abbasvand2021,Abbasvand2020a,Abbasvand2020b} with the Interface Region Imaging Spectrograph (IRIS) found that the acoustic energy flux is sufficient to balance the radiative losses in the quiet chromosphere, though its contribution to the radiative losses in the active-region chromosphere is only 10--30 per cent.

In short, it is widely accepted that acoustic waves, as the fundamental energy transportation and dissipation mechanism, are enough to heat and sustain the solar chromosphere in the quiet regions. Turning to the coronal heating problem, the wave periods observed in the corona exhibit a peak of around three to five minutes~\cite{DeMoortel2002, VanDoorsselaere2008, Tomczyk2009, Morton2016, Morton2019}, which is likely connected with photospheric $p$-mode oscillations with the same peak period. However, the chromosphere is an acoustic cut-off layer, in which waves with periods around, or longer than, {three} minutes are evanescent and non-propagating. Thus, coronal heating cannot be directly maintained by acoustic waves from the photosphere, and it is widely believed that the magnetic field plays an essential role in the transportation of energy from the photosphere to the corona~\cite{Abbett2007}. At present, almost all possible magnetic field effects, such as Alfv\'{e}nic waves (e.g.,~\cite{Dmitruk2002,Li2003,Sokolov2021}), fast/slow magneto-acoustic waves (e.g.,~\cite{Bel1977,Pontieu2004,Julia2021}), and magnetic reconnections (e.g.,~\cite{Vekstein1994, Pueschel2014}), have been applied to the explanation of coronal heating, which is closely related to various magnetic phenomena observed by ground and space instruments with higher and higher spatial and temporal resolution~\cite{Narain1996, Aschwanden2005, Erdelyi2007}.

Although the importance of the magnetic field has become the consensus in solar physics, some observational research has revealed that the contribution of acoustic waves cannot be simply and arbitrarily ignored in corona, for example, {the work of Morton et al. (2019)~\cite{Morton2019}}. They measured the Alfv\'enic wave motions along line-of-sight through the Doppler shift of the Fe XIII emission line above the limb in corona and found that there was always a peak about $4\ \rm{mHz}$ appearing in the power spectrum of the Doppler velocity time-series taken from the iron line observation. They finally confirmed the ubiquity of $4\ \rm{mHz}$ in corona with arbitrary power spectral density (PSD) measurements using the data from 2005 to 2015, which cover different phases of the whole 11-year solar cycle. In order to explain these observational results, Morton et al. (2019)~\cite{Morton2019} suggested magnetohydrodynamical (MHD) wave models where coronal Alfv\'enic modes are excited at the transition region {(TR)} through the acoustic-Alfv\'enic mode conversion~\cite{Cally2008,Cally2011,Cally2017}, and the inputting acoustic waves are leaked into the atmosphere from the photosphere by magneto-acoustic portals~\cite{Jefferies2006}.

However, the point of this suggestion that the inputting acoustic waves are leaked from the photosphere may conflict with the observational ubiquity of $4\ \rm{mHz}$, as a chromosphere leaking acoustic waves everywhere is equivalent to being transparent to acoustic waves. Thus, it is necessary to explore the origin of the global $4\ \rm{mHz}$ oscillations in corona and investigate the possibility of coronal heating via the resulting shock waves. For this goal, we performed global spherically symmetric {hydrodynamical (HD)} simulations of the solar atmosphere between $1$ and $1.25\ R_{\odot}$. For simplicity, we applied an artificial energy flux in the chromosphere, with which the simulated chromosphere could quickly converge to a quasi-steady state that agreed with various observations. The computation time was thus greatly reduced, enabling us to explore the global structure of the solar atmosphere. This paper is arranged as follows: The model is presented in Section \ref{Sec:Model}. The numerical settings, such as the computational grid, initial conditions, and boundary conditions, are presented in Section \ref{Sec:settings}. The numerical results are presented and discussed in Section \ref{Sec:results}. We summarize and conclude in Section \ref{Sec:conclusions}.

%%%%%%%%%%%%%%%%%%%%%%%%%%%%%%%%%%%%%%%%%%
\section{The Model}\label{Sec:Model}
%=========================================
\subsection{Basic Equations}

We consider a spherically symmetric HD solar atmosphere. The following are the basic equations adopted in our model, including the mass conservation,
\begin{equation}\label{eq_mass}
\frac{\partial\rho}{\partial t}+\frac{1}{r^2}\frac{\partial}{\partial r}\left(r^2\rho v_r\right)=0,
\end{equation}
radial momentum conservation,
\begin{equation}\label{eq_momentum}
\frac{\partial(\rho v_r)}{\partial t}+\frac{1}{r^2}\frac{\partial}{\partial r}\left[r^2\left(\rho v^2_r+p\right)\right]=-\frac{GM_{\odot}}{r^2},
\end{equation}
and total energy (including internal energy and kinetic energy) conservation equations,
\begin{equation}\label{eq_energy}
\frac{\partial e}{\partial t}+\frac{1}{r^2}\frac{\partial}{\partial r}\left[r^2v_r\left(e+p)\right)\right]=-\rho v_r\frac{GM_{\odot}}{r^2}+q^-+\frac{\partial F_{\rm c}}{\partial r},
\end{equation}
where $t$, $r$, $\rho$, $v_r$, $p$, $e$, $G$, $M_{\odot}$, $q^-$, and $F_{\rm c}$ are the time, spherical radius, density, radial velocity, pressure, total energy, gravitational constant, solar mass, cooling rate, and artificial chromospheric energy flux, respectively {(we ignore the heat conduction in our model, though it is important, especially within the TR \endnote{{There are two reasons for this omission. One is that the resolution of our model is too low to accurately reveal the TR. Another is that the tiny time-step due to the diffusion calculation for the heat conduction is unfavorable for long-term physical time integration. According to our computational practice, the effect of ignoring heat conduction does not alter the generation of $4\ \rm{mHz}$ oscillations, which appear in the early stages of our simulations, but the code of the heat conduction is too slow to reveal its long-term effects through sufficient long time-integration.}})}. We treat the plasma as an ideal gas, so that $e=p/(\gamma-1)+\rho v^2_r/2$, where $\gamma=5/3$ is the adiabatic index. The numerical simulations were performed with Athena++~\cite{Athena++} {(Specifically, we used its default HD module including the HLLC Riemann solver, time-integrator of second-order accurate van Leer predictor--corrector scheme, and the spatial reconstruction of piecewise linear method  applied to primitive variables)}, and accordingly Equations \eqref{eq_mass}--\eqref{eq_energy} are written in numerical conservation forms, and the effects of gravity, radiative cooling, and artificial energy flux are all treated as physical source terms appearing at the right-hand side of the equations.

%=========================================
\subsection{Artificial Chromospheric Energy Flux}\label{Sec:ch_flux}

\begin{figure}[H]
%	\centering
	\includegraphics[width=1.0\linewidth]{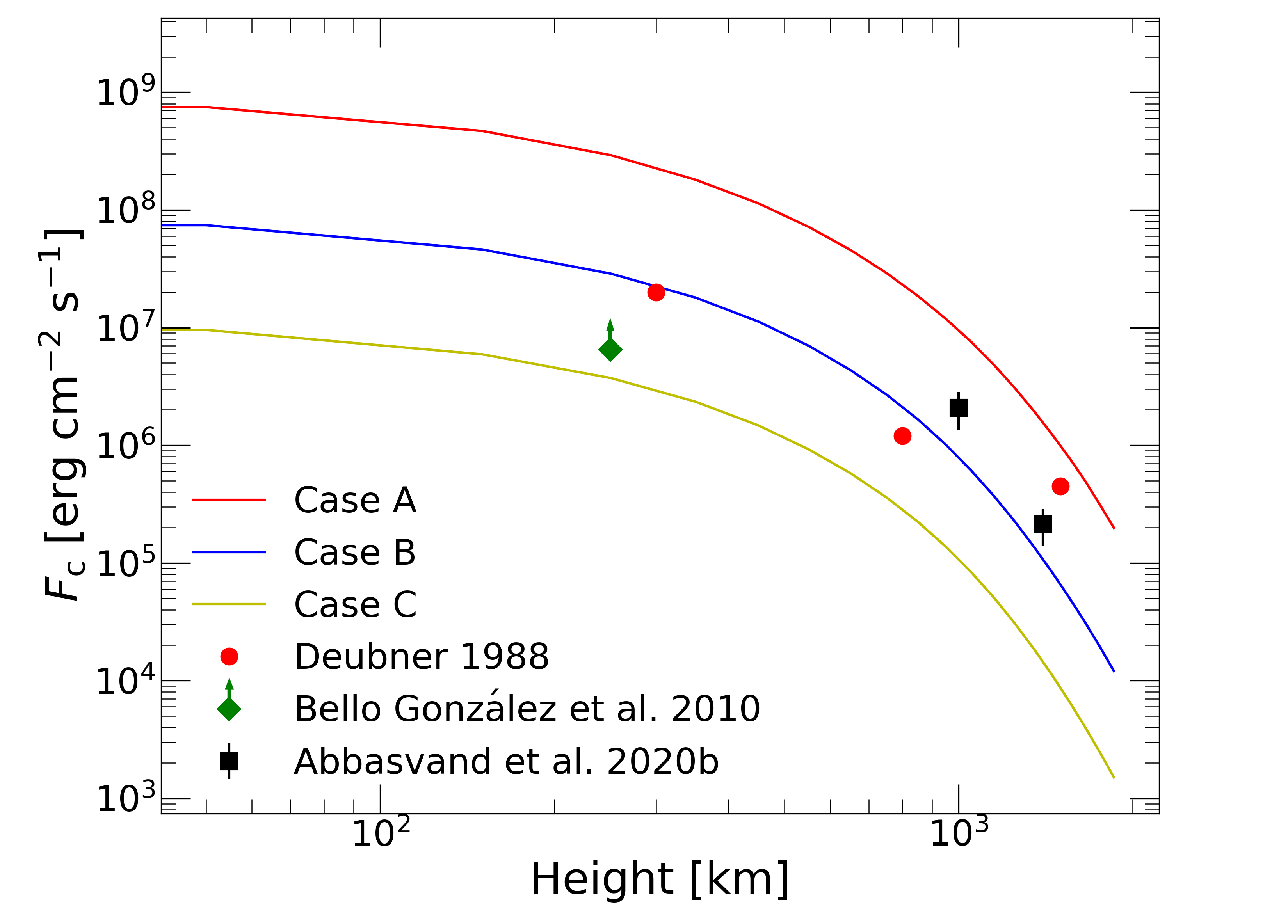}
	\caption{Artificial %MDPI: We moved Figure 1 after its first citation, please check.
 chromospheric energy fluxes $F_{\rm{c}}$ (solid lines) and the observed acoustic energy fluxes (dots, diamonds, and squares). The curves of $F_{\rm{c}}$ (only defined in the height range 0--2000 km) were plotted with the mean pressures averaged over the last 100 days of the respective simulations. The dots, diamonds, and squares correspond to the observed acoustic energy fluxes in the solar chromosphere, obtained from~\cite{Deubner1988,Bello2010,Abbasvand2020b}, respectively. The arrow in the diamond signifies that the data from~\cite{Bello2010} are the lower limit, {and Deubner (1988)~\cite{Deubner1988} did not provide error information for the three dots.}}
	\label{fig:artificial_flux}
\end{figure}

As mentioned in Section \ref{intro}, it is widely accepted that the main heating mechanism in the quiet chromosphere is the dissipation of acoustic waves transmitted from the photosphere. Ideally, we should set an acoustic wave generator in the photosphere as the lower boundary condition (e.g., ref.~\cite{Carlsson1992} used a sinusoidal pistol as the lower boundary). However, as explained below, we instead use an artificial chromospheric heating flux in the chromosphere of our simulation (between $0$ and $2000$ km above the photosphere), which is
\begin{equation}\label{eq_aflux}
F_{\rm c}=k_0 p,
\end{equation}
where $k_0$ is an coefficient whose value will be discussed in Section \ref{Sec:results}, and the photosphere boundary is set to be in hydrostatic equilibrium (see Section \ref{BdryConds} for more details). The reason we use this artificial flux is as follows. First, our main finding in this paper, as will be discussed in Section \ref{Sec:results}, is that acoustic shock waves are newly generated at the bottom of the corona, which propagate upward and participate in the coronal heating. Thus, we need to exclude the possibility that these acoustic waves are leaked from the photosphere, and a boundary condition with the acoustic wave input from the photosphere should be avoided. Secondly, we focused our computation on the corona, so that we only had very limited resolution in the chromosphere ($\sim$20 cells). Had we used the real heating flux via the dissipation of acoustic waves and the corresponding lower boundary condition, this low resolution would make it more difficult to converge to a quasi-steady solution in the chromosphere, where the heating from the dissipation of acoustic waves must be calculated precisely, and make the acoustic waves from the photosphere more likely to leak through the chromosphere. Thus, here, we used an artificial chromospheric heating flux instead. We found that the simulated chromosphere could quickly converge to a quasi-steady state which conformed to various observations (see Section \ref{modelat}) with this artificial flux and limited resolution in computation, while avoiding the possible leakage of acoustic waves from the photosphere. Note that this flux was only applied in the chromosphere. Considering that the observed thickness of the solar chromosphere is $\sim$2000 km, we only added this energy flux in the height range between $0$ and $2000$ km above the photosphere ($r=R_{\odot}$), which was the inner boundary of our calculation. {Although the real chromosphere energy flux may not suddenly stop at $2000$ km, the pressure there becomes relatively small due to the rapid decrease in density (but the temperature change is not significant in chromosphere), and hence the artificial energy flux, which is proportional to the pressure, also becomes very small and the actual impact of this truncation should not be significant. Meanwhile, this approach of fixing the truncation position at a reliable observed chromospheric thickness, rather than releasing it as a free parameter, simplifies the model and ultimately concentrates any uncertainty in a single parameter $k_0$, making it easier to fit the other observations (see Figure \ref{fig:artificial_flux}).}

Physically speaking, the energy flux of an acoustic wave is
\begin{equation}\label{eq_flux}
F=\frac{1}{2} \rho v_\mathrm{a}^2 c_\mathrm{s}
\end{equation}
where $v_\mathrm{a}$ is the velocity amplitude and $c_\mathrm{s}$ is the sound speed. The pressure is dominated by gas pressure, so that $p=\rho k T/(\mu m_\mathrm{p})$, where $k$ is the Boltzmann constant, $\mu$ is the mean molecular weight, and $m_\mathrm{p}$ is the proton mass. At different heights of the chromosphere, the variation in density is much larger than that of the temperature, so that we have approximately $\rho \propto p$, and $c_\mathrm{s}$ can also be regarded as a constant. Thus, we have $F \propto p$ when the variation in $v_\mathrm{a}$ is neglected. Anderson et al. (1989)~\cite{Anderson1989a,Anderson1989b} (also see the review of Narain and Ulmschneider (1996)~\cite{Narain1996}) found that the solar chromospheric cooling flux is for many scale heights proportional to the gas pressure, which is consistent with the directly observed propagating acoustic wave flux~\cite{Endler1983,Deubner1988}. Anderson et al. (1989)~\cite{Anderson1989a} attributed this phenomenon to the saturation of the velocity amplitude of acoustic waves near the speed of sound. So, eventually we could approximately obtain $F \propto p$, which is the formula for the artificial chromospheric heating flux we used. The parameter $k_0$ can be determined from observations (see Section \ref{Sec:results}).

{However, it is worth noting that this artificial energy flux is highly simplified, ignoring many details of chromosphere heating and only applicable to our simplified model. For further studies in the future, it will be necessary to refer to detailed chromosphere heating studies (e.g.,~\cite{Zhang2021}).}

%=========================================
\subsection{Radiative Cooling}
We assume that the low-density plasma in the chromosphere and corona is in the collisional ionization equilibrium (CIE) everywhere and ignore radiative transfer. Therefore, the radiative cooling rate can be written as
\begin{equation}\label{eq_coolingrate1}
q^-=n_{\rm e}n_{\rm i}\Lambda_1,
\end{equation} 
where $n_{\rm e}$ and $n_{\rm i}$ are the number densities {(in $\rm{cm^{-3}}$)} of electrons and ions, respectively, and {$\Lambda_1$} is the normalized cooling rate (in $\rm{erg\ cm^3\  s^{-1}}$). Sutherland and Dopita (1993)~\cite{CoolingRate} calculated {$\Lambda_1$} under CIE, including significant collisional line radiation, continuum radiation, and recombination heating. They presented a series of models with different metal abundances covering the temperature range of $10^4$--$10^{8.5}\ \rm K$. We adopted the model for solar abundances in this paper. 

For the cooling of plasma with solar abundances at temperatures below $10^4$ K, we adopted the formula of Mashchenko et al. (2008)~\cite{Mashchenko2008},
\begin{eqnarray}
&&q^-=n^2_{\rm H}\Lambda_2,  \label{eq_coolingrate2}\\
&&\log\left(\Lambda_2\right)=-24.81+2.928x-0.6982x^2, \label{def_Lambda2}
\end{eqnarray}
where $x\equiv\log(\log(\log(T)))$, $T$ is temperature (in K), $n_{\rm H}$ is the density number of hydrogen atoms (in $\rm{cm^{-3}}$), {and the normalized cooling rate $\Lambda_2$ is defined in Equation \eqref{def_Lambda2}.} This formula is valid from $20$ to $10^4$ K, but we had to stop using it below $5000$ K to avoid over-cooling in the simulations. Therefore, the plasma below $5000$ K is adiabatic (which is still able to achieve a temperature below $5000$ K via adiabatic expansion). 

%=========================================
\subsection{State of Matter}
In the solar atmosphere, the temperature of plasma rises from thousands to millions of Kelvins, and the state of matter can be described by a partially ionized ideal gas. We assumed that the plasma reached local thermal equilibrium (LTE) everywhere, which is compatible with the CIE assumption. As radiative transfer was neglected in our calculation, the radiation pressure was consequently also ignored under LTE, so that the pressure of plasma can be calculated as
\begin{eqnarray}
&&p = \left(n_{\rm i}+n_{\rm e}\right)kT,\label{eq_pressure}\\
&&n_{\rm i} = \frac{\mathcal{R}\rho}{\mu_{\rm i}},\\
&&n_{\rm e} = E n_{\rm i},
\end{eqnarray}
where $k$, $\mathcal{R}$, $\mu_{\rm i}$, $E$, and $T$ are the Boltzmann constant, gas constant, mean molecular weight, average number of electrons lost per atom, and temperature of the plasma, respectively. We regarded the plasma as a mixture of hydrogen and helium, and ignored the contribution of trace amounts of metal elements, so that $E$ can be calculated as
\begin{eqnarray}
&& E = Y_{\rm H}\eta\sss[\rm H]+Y_{\rm{He}}\left(\eta\sss[\rm{He}]+2\eta\sss[\rm{He^+}]\right),\label{eq_E}\\
&& Y_{\rm H} = \frac{4X_{\rm H}}{4X_{\rm H}+X_{\rm{He}}},\\
&& Y_{\rm{He}} = \frac{X_{\rm{He}}}{4X_{\rm H}+X_{\rm{He}}},
\end{eqnarray}
where $Y_{\rm H}$ and $Y_{\rm{He}}$ are the number fractions of hydrogen and helium; $X_{\rm H}$ and $X_{\rm{He}}$ are their mass fractions; and $\eta\sss[\rm H]$, $\eta\sss[\rm{He}]$, and $\eta\sss[\rm{He^+}]$ are the ionization fractions of hydrogen and the first- and second-order ionized helium, respectively. Under both LTE and CIE, these ionization fractions satisfy Saha's equations~\cite{Saha1920}
\begin{eqnarray}
\frac{\eta\sss[\rm H]}{1-\eta\sss[\rm H]}\frac{E}{1+E} =\frac{(2\pi m_{\rm e})^{3/2}(kT^{5/2})}{p h^3} e^{-\chi\sss[\rm H]/kT},\label{eq_saha1}&\\
\frac{\eta\sss[\rm{He}]}{1-\eta\sss[\rm{He}]-\eta\sss[\rm{He^+}]} \frac{E}{1+E} = 4\frac{(2\pi m_{\rm e})^{3/2}(kT^{5/2})}{p h^3} e^{-\chi\sss[\rm He]/kT},&\\
\frac{\eta\sss[\rm{He^+}]}{\eta\sss[\rm{He}]}\frac{E}{1+E} = \frac{(2\pi m_{\rm e})^{3/2}(kT^{5/2})}{p h^3} e^{-\chi\sss[\rm{He^+}]/kT},\label{eq_saha3}&
\end{eqnarray}
where $m_{\rm e}$, $h$, $\chi\sss[\rm H]$, $\chi\sss[\rm{He}]$, and $\chi\sss[\rm{He^+}]$ are the electron mass, Plank constant, and ionization energies of the corresponding elements, respectively.

In simulations, the density and pressure are known in the cells of the numerical grid. To calculate the cooling rate of each cell with Equations (\ref{eq_coolingrate1}) and (\ref{eq_coolingrate2}), the key is to evaluate $E$, which can be achieved by solving Equations (\ref{eq_E}) and  (\ref{eq_saha1})--(\ref{eq_saha3}) for unknown quantities $E$, $\eta\sss[\rm H]$, $\eta\sss[\rm{He}]$, and $\eta\sss[\rm{He^+}]$. In practice, we prepared a 2D table of $E$ for different possible cases of density and pressure through solving these equations with a non-linear system solver. With this table, we could evaluate $E$ with a bilinear interpolation in the computation.

{It is worth noting that the real chromosphere is clearly in a non-equilibrium ionization state, in which the ionization/recombination timescale of hydrogen is obviously longer than the HD/MHD timescale~\cite{Carlsson2002}. This would make the chromospheric plasma exhibit larger temperature fluctuations than those predicted from LTE when the shock wave sweeps it~\cite{Martinez2020}. However, in our model, the chromosphere is highly simplified, and a lot of real details are ignored, including the acoustic shock from the photosphere. Therefore, we assumed that the impact of LTE on our quasi-hydrostatic chromosphere structure is limited, which can be confirmed by our subsequent numerical results (see Section \ref{modelat}).}

%%%%%%%%%%%%%%%%%%%%%%%%%%%%%%%%%%%%%%%%%%
\section{Numerical Settings}\label{Sec:settings}
%=========================================
\subsection{Grid and Initial Conditions}
A logarithmically equally spaced grid with $1550$ cells was adopted in our simulations from the top of the photosphere $r=R_{\odot}$ to $r=1.25R_{\odot}$. This grid had a minimum spacing of $\sim$100 km at $r=R_{\odot}$, and the chromosphere (thickness of $\sim$2000 km) could be resolved with $\sim$20 cells. Each simulation run was carried out for long enough to achieve a quasi-steady state. Through repeated attempts, we found that the final result was not sensitive to the initial conditions {or the grid resolution}. Therefore, we adopted a low-density and low-temperature static homogeneous plasma as the initial condition in our simulations to minimize the computational consumption. Specifically, the plasma in the grid was uniformly initialized to $\rho=2\times10^{-14}\ \rm{g\ cm^{-3}}$, $T=10^4\ \rm{K}$, and $v_r=0\ \rm{cm\ s^{-1}}$.

%=========================================
\subsection{Boundary Conditions}\label{BdryConds}
There are two boundaries in our model. One is the outer boundary at $r=1.25R_{\odot}$, which is far below the solar wind acceleration region ($r\gtrsim3R_{\odot}$, see~\cite{Aschwanden2005}), and the other is the photosphere boundary at $r=R_{\odot}$. The outer boundary was set as an unidirectional outflow boundary through a simple improvement of the classical outflow conditions, in order to prevent fictitious inflow from outside and ensure normal outflow. {Specifically, we inserted a function into the integrator loop to monitor the direction of the mass flux at the interface between each boundary cell and the corresponding ghost cell. If fictitious mass inflow from the ghost cell was detected, all of the fluxes, including the mass, momentum, and energy fluxes, were reset to zero before they were used to update the cell-centered conservative variables in the boundary cell.} This kind of unidirectional boundary was applied in our previous work~\cite{Xue2021}, where its validity was confirmed.

The photosphere boundary plays an important role in the formation of the solar atmosphere. Since we adopted a second-order-accuracy algorithm in our simulations, there were two ghost cells at the boundary. The centers of these two cells were set inside the photosphere at 50 km and 150 km below $r=R_{\odot}$, respectively. The cells were so thin that the density difference was insignificant, as observed in the photosphere (see the density profile in~\cite{Vernazza1981}). Therefore, we assumed that the densities in both ghost cells were the same as that of the photosphere top, which was $\sim$$2\times10^{-7}\ \rm{g\ cm^{-3}}$ from observations. We also assumed hydrostatic equilibrium in these two ghost cells. Note that the observed photosphere was actually not in hydrostatic equilibrium, and our assumption led to no self-consistent energy flow being generated from this boundary to heat the chromosphere. An artificial chromospheric energy flux (see Section \ref{Sec:ch_flux}) was adopted to compensate for this inconsistency, while avoiding unnecessary computational complexity.

%%%%%%%%%%%%%%%%%%%%%%%%%%%%%%%%%%%%%%%%%%
\section{Numerical Results}\label{Sec:results}
%=========================================
\subsection{Structure of the Model Atmosphere}\label{modelat}

In our model, there is only one parameter, the ratio of the artificial chromospheric energy flux to the local pressure, $k_0$. Multiple simulation runs with different values of $k_0$ were performed, and we found that the larger the value of $k_0$, the higher the mean temperature and density of the corona in the quasi-steady state became. This is to be expected, because when $k_0$ increases, the heating flux in the chromosphere increases and consequently more material with higher temperature is driven into the corona. We also found that there existed a critical lower limit of $k_{0c}=2.01\times10^2\rm{cm\ s^{-1}}$, below which the model calculation collapsed due to lack of heating to sustain the chromosphere and subsequently the corona. In order to reveal the effects of $k_0$, here, we present the simulation results for three different values of $k_0$ (Table \ref{tab:3cases}), i.e., Cases A, B, and C, with $k_0= 1.61 \times 10^4$, $1.61 \times 10^3$, and $2.1 \times 10^2$ cm s$^{-1}$, respectively. Case B was actually the optimal simulation result that we found (see discussion below). Case C corresponded to a value of $k_0$ slightly greater than $k_{0c}$, which can be seen as a lower limit. Case A represents the simulation result with a larger value of $k_0$, presented as a comparison. Each simulation was carried out for one year (365 days) in physical time, which achieved a quasi-steady state after about 50 days {(Our code is serial and does not require much memory from the computer, but due to the high temperature and low density of our research object, coronal plasma, the practical computational time-step is small, so the entire computation of each simulation took about 2 days of CPU time)}.

\begin{table}[H]
	\caption{Important characteristics for three typical cases of simulation. The $p$-mode frequencies indicate the range of peak frequencies in the velocity PSD measured at different heights. The mass losses are the mean mass outflow rates at the outer boundary of our model. The max temperatures are maximal mean temperatures averaged over the last 100 days of the respective simulations.}
	\label{tab:3cases}
	\newcolumntype{C}{>{\centering\arraybackslash}X}
	\begin{tabularx}{\textwidth}{cCCCc}
		\toprule
		\textbf{Case} & \boldmath{$k_0$} & \textbf{\boldmath{$p$}-Mode Freq.} & \textbf{Mass Loss} & \textbf{Max Temp.}\\
		& \boldmath{$[\rm{\times 10^3 cm\ s^{-1}}]$} & \textbf{[mHz]} & \textbf{[\boldmath{$\rm{\times 10^{-11}g\ cm^{-2}\ s^{-1}}$}]} & \textbf{[\boldmath{$\rm{\times 10^6K}$}]}\\
		\midrule
		A & 16.1 & 3.39--4.98 & 6.8 & 1.1\\
		B & 1.61 & 3.90--4.07 & 1.4 & 0.92\\
		C & 0.21 & 3.99--4.38 & 0.20 & 0.84\\  
		\bottomrule
	\end{tabularx}	
\end{table}

The value of $k_0$ can be constrained by observations. Several works in literature presented the acoustic energy fluxes obtained from observations, which can be compared directly with those from our simulation results, as presented in Figure \ref{fig:artificial_flux}. Deubner (1988)~\cite{Deubner1988} found the acoustic energy fluxes to be $2.0\times10^7$, $1.2\times10^6$, and $4.5\times10^5$ $\rm{erg\ cm^{-2}\ s^{-1}}$ at the heights $300$, $800$, and $1500$ km (also see~\cite{Narain1996}), respectively, represented by the red dots in Figure \ref{fig:artificial_flux}. Bello et al. (2010)~\cite{Bello2010} proposed an average acoustic energy flux at $250$ km of $6.5\times10^6$ $\rm{erg\ cm^{-2}\ s^{-1}}$ as the lower limit, represented by the green diamonds. Abbasvand et al. (2020)~\cite{Abbasvand2020b} presented acoustic energy fluxes at the heights of $1000$ and $1400$ km, which were $2.085 \pm 0.744\times10^6$ and $2.14 \pm 0.74\times10^5$ $\rm{erg\ cm^{-2}\ s^{-1}}$, respectively, in the non-magnetic area, represented by the black squares. The solid lines represent the artificial chromospheric energy fluxes $F_{\rm{c}}$ for the three different cases, with the red line for Case A, blue line for Case B, and yellow line for Case C. We can see that the observed fluxes all fall between the profiles of Case A and Case C, while Case B best conformed to the observations.

Another constraint was the mass loss rate of the solar atmosphere in our simulation, which was the mass outflow rate at the outer boundary. Cohen (2011)~\cite{Cohen2011} calculated the solar mass loss rate from observations, which was scattered around the value of $2\times 10^{-14}$ M$_\odot$ yr$^{-1}$ with a variation factor of 2--5. The center value corresponded to $1.3\times 10^{-11}\rm{g\ cm^{-2}\ s^{-1}}$ at the outer boundary (1.25 $R_\odot$) of our simulation. Meanwhile, the mean mass loss rates (see Table \ref{tab:3cases}) were $6.8\times10^{-11}\rm{g\ cm^{-2}\ s^{-1}}$ for Case A,  $1.4\times10^{-11}\rm{g\ cm^{-2}\ s^{-1}}$ for Case B, and $0.20\times10^{-11}\rm{g\ cm^{-2}\ s^{-1}}$ for Case C, averaged over the last 100 days of the respective simulations. We can see that Case B also agreed quite well with the observed mass loss rate.

\begin{figure}[H]
%	\centering
	\includegraphics[width=1.0\linewidth]{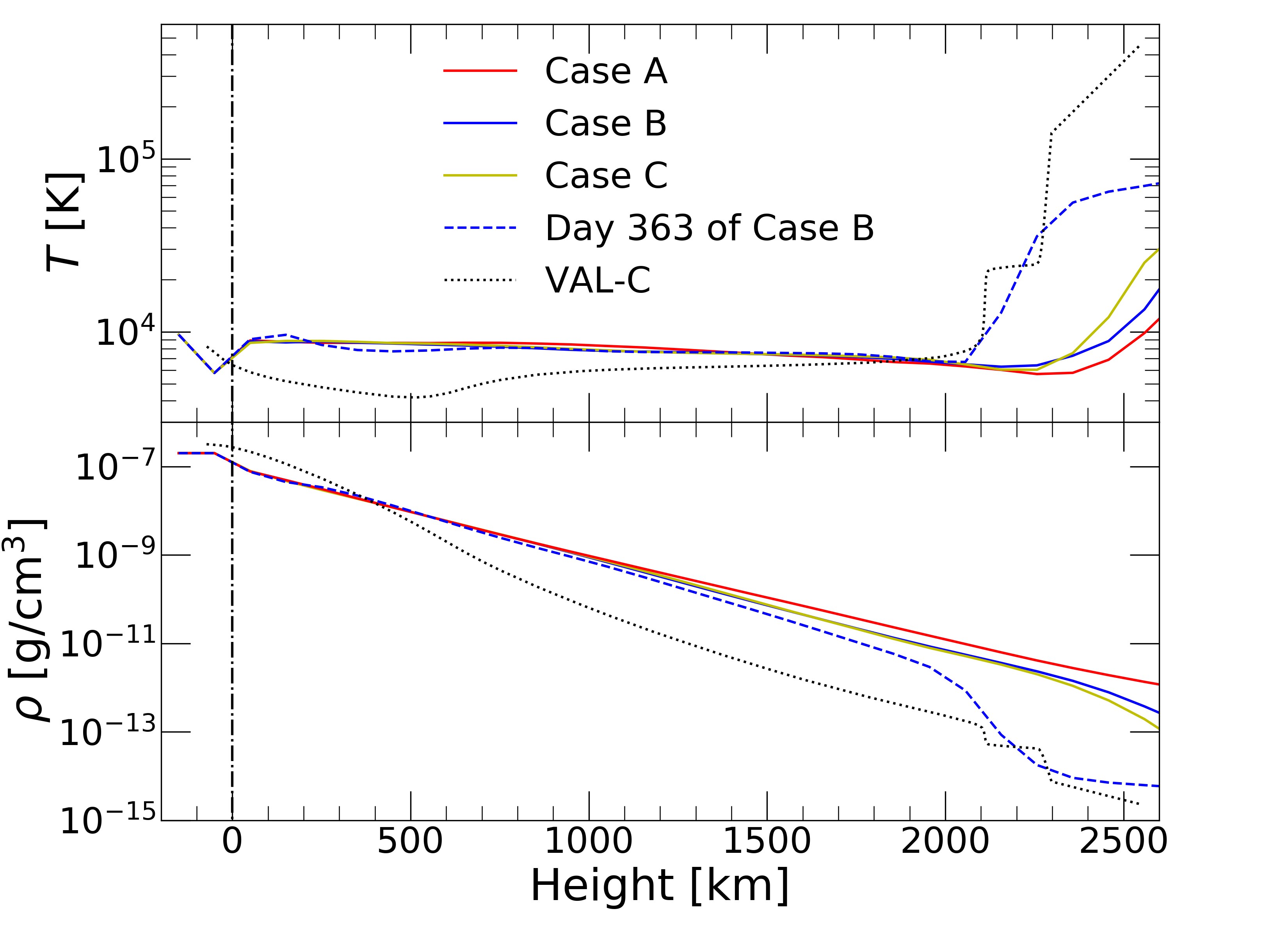}
	\caption{Mean temperature and density profiles for the three cases (solid lines), the temperature and density profiles on Day 363 of the Case B simulation (dashed lines), and the corresponding profiles in the VAL-C model (dotted lines). The vertical dot-dashed line denotes the top of the solar photosphere ($r=R_{\odot}$).}
	\label{fig:num_vs_obs}
\end{figure}

In Figure \ref{fig:num_vs_obs}, we compare the structure of the simulated solar atmosphere with the VAL-C model determined from \textit{Skylab} observations of the quiet Sun~\cite{Vernazza1981}. The solid lines with different colors represent the mean density and temperature profiles of the simulation results, averaged over the last 100 days of the respective simulations. The dotted lines represent the density and temperature profiles of the VAL-C model. In each simulation, there were some time points when the density and temperature profiles agreed better with those of the VAL-C model. For clarity, here, we only show the profiles on Day 363 of Case B, represented by the dashed lines. While the exact values differed slightly from the VAL-C model, the tendency of the variation and the location of {TR} agreed well. The tendency of variation in the mean density and temperature profiles (solid lines) also agreed with those of the VAL-C model (dotted lines), though the location of the {TR} differed slightly. Considering that we have very limited resolution for the chromosphere ($\sim$20 cells), we think that the agreement is acceptable and the main characteristics of the solar chromosphere were reproduced by our simulations, especially Case B, which also agreed quite well with the other observations, as mentioned above.

{It is worth noting that the detailed deviation between our model and the observational case is also obviously shown in Figure \ref{fig:num_vs_obs}, though the tendency is the same between them. From the temperature profiles shown in the upper panel, it can be seen that the model temperature at low chromosphere was higher than the real case but it reversed after stopping the artificial energy flux at 2000 km. This reflects the drawback of our model on the estimation of the chromospheric heating--cooling balance. This results from the over simplification of our model chromosphere, which could be removed by further consideration of real chromospheric details in future works.}

\begin{figure}[H]
%	\centering
	\includegraphics[width=1.0\linewidth]{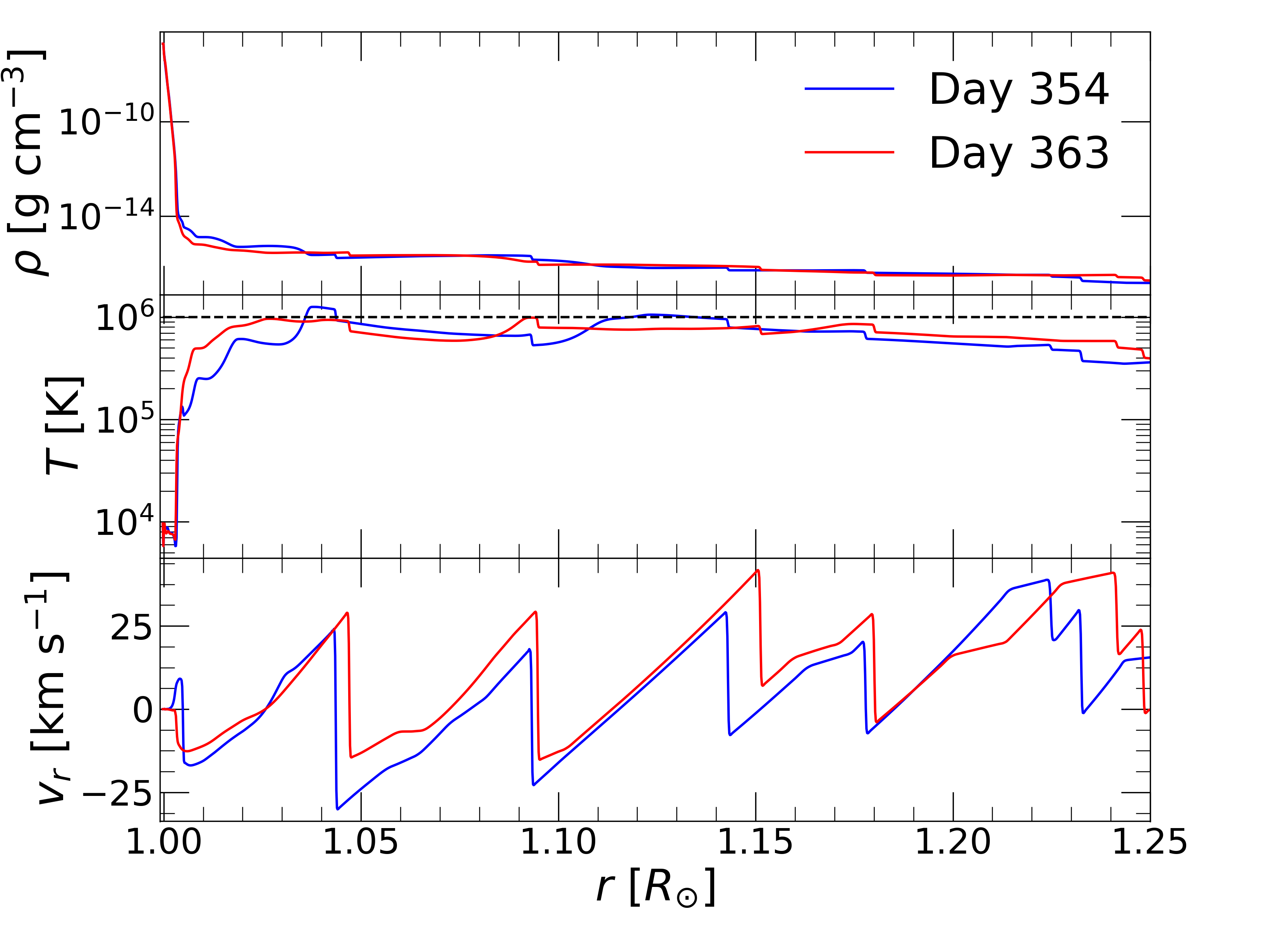}
	\caption{Density, temperature, and radial velocity profiles on Days 354 (blue lines) and 363 (red lines) of the Case B simulation. The horizontal dashed line in the middle panel denotes the temperature of $10^6$K.  }
	\label{fig:rho_T_Vr_profiles}
\end{figure}

In Figure \ref{fig:rho_T_Vr_profiles}, we present the density, temperature, and radial velocity profiles on Days 354 and 363 of Case B, which exhibit two typical states of the quasi-steady atmosphere. It can be seen that a series of shocks were generated at the bottom of the corona and propagated upward {with a speed $\sim$$139\ \rm{km\cdot s^{-1}}$, which could be estimated through multiplying the spatial distance between shocks ($\sim$$0.05 R_{\odot}$, see the radial velocity profile shown at the bottom panel of Figure \ref{fig:rho_T_Vr_profiles}) by their passing frequency ($4\ \rm{mHz}$, see Section \ref{shockwave}).} These traveling shocks heated the coronal plasma, since the temperature post shock is always higher than that of the shock front (see the middle panel). While these two states $9$ days apart had similar profiles in general, a higher density always resulted in a cooler plasma through a larger radiative cooling under almost the same shock heating (almost the same shock strength, see the bottom panel). The maximum temperatures on these two days both reached $10^6$ K.

\begin{figure}[H]
%	\centering
	\includegraphics[width=1.0\linewidth]{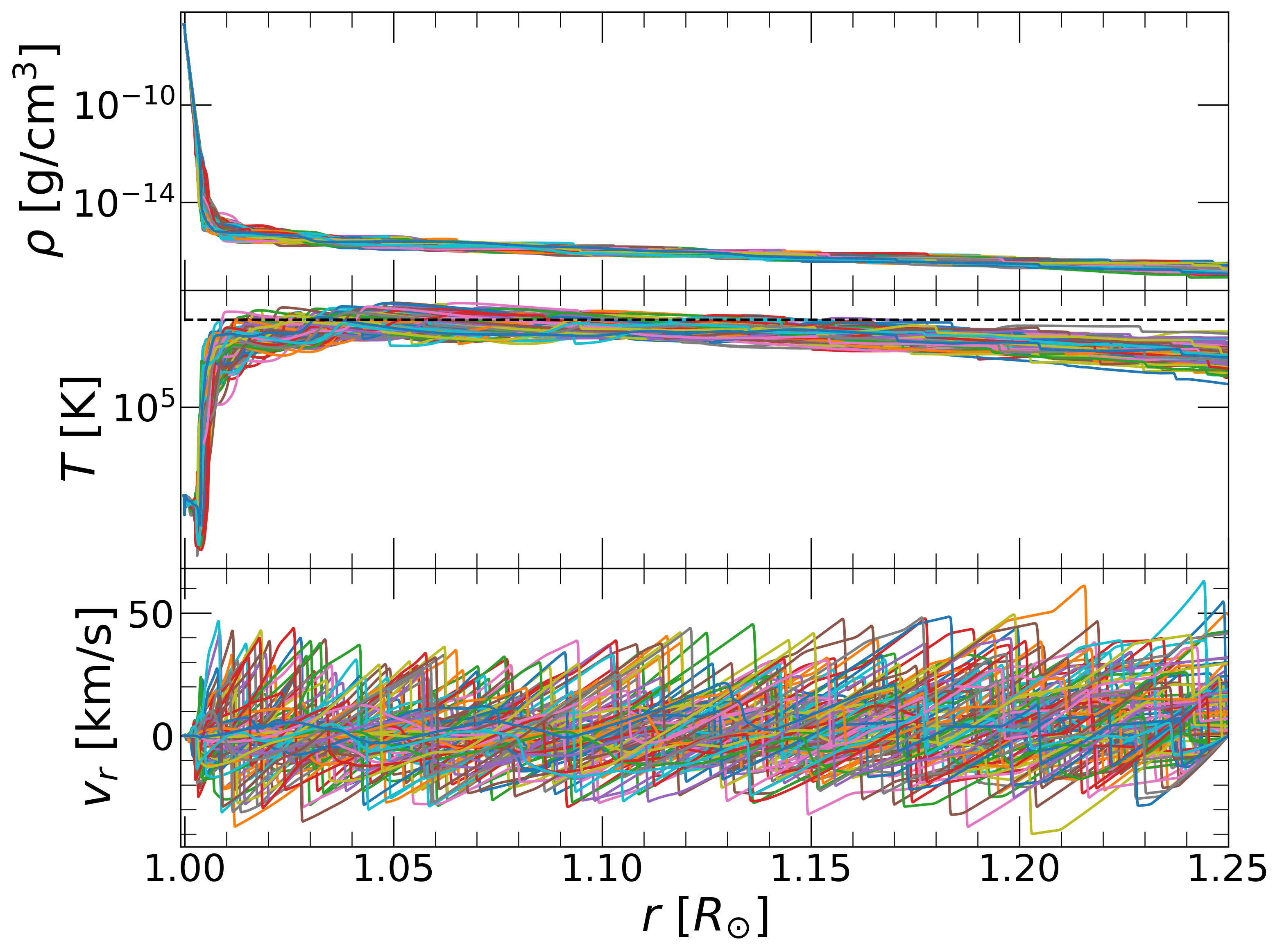}
	\caption{Density, %MDPI: Please confirm if the explanation of the colors needs to be added in the figure caption.
 temperature, and radial velocity profiles over the last 100 days (from Day 265 to Day 365) of the Case B simulation.}
	\label{fig:last_50_days}
\end{figure}

In order to show fluctuations in the quasi-steady model atmosphere, we drew the density, temperature, and radial velocity profiles of the last 100 days (from Day 265 to Day 365) in Figure \ref{fig:last_50_days}. While there was little variance in the model chromosphere ($r<1.005R_{\odot}$), obvious fluctuations occurred in the model corona ($r \geq 1.005R_{\odot}$), with a density ranging from $\sim$$10^{-18}$ to $\sim$$10^{-15}\ \rm{g\ cm^{-3}}$, temperature from $\sim$$10^5$ to $\sim$$2\times10^6$ K, and radial velocity from $\sim$$-40$ to $\sim$$40\ \rm{km\ s^{-1}}$, respectively. The amplitude of the coronal heating shocks in our model was much larger than that required by chromospheric heating, whose typical radial velocity jump is only a few kilometers per second~\cite{Zhang2021}, while the density in the corona was much smaller than that in the chromosphere, so it is natural that the model corona could be heated to a much higher temperature. It should be noted that, in observations, the amplitude of acoustic shocks in the corona is not so large (e.g.,~\cite{Doorsselaere2020} and references therein). This is because our simulation was hydrodynamic and did not include the effects of magnetic fields. When magnetic fields are added, the energy stored in acoustic waves can be converted into other forms, such as Alfv{\'e}nic waves (see Section \ref{shockwave}), spicules (see Section \ref{shockwave}), and prominences (see Section \ref{upflows}), which explains why such high amplitude acoustic waves are not typically observed.

\begin{figure}[H]
%	\centering
	\includegraphics[width=1.0\linewidth]{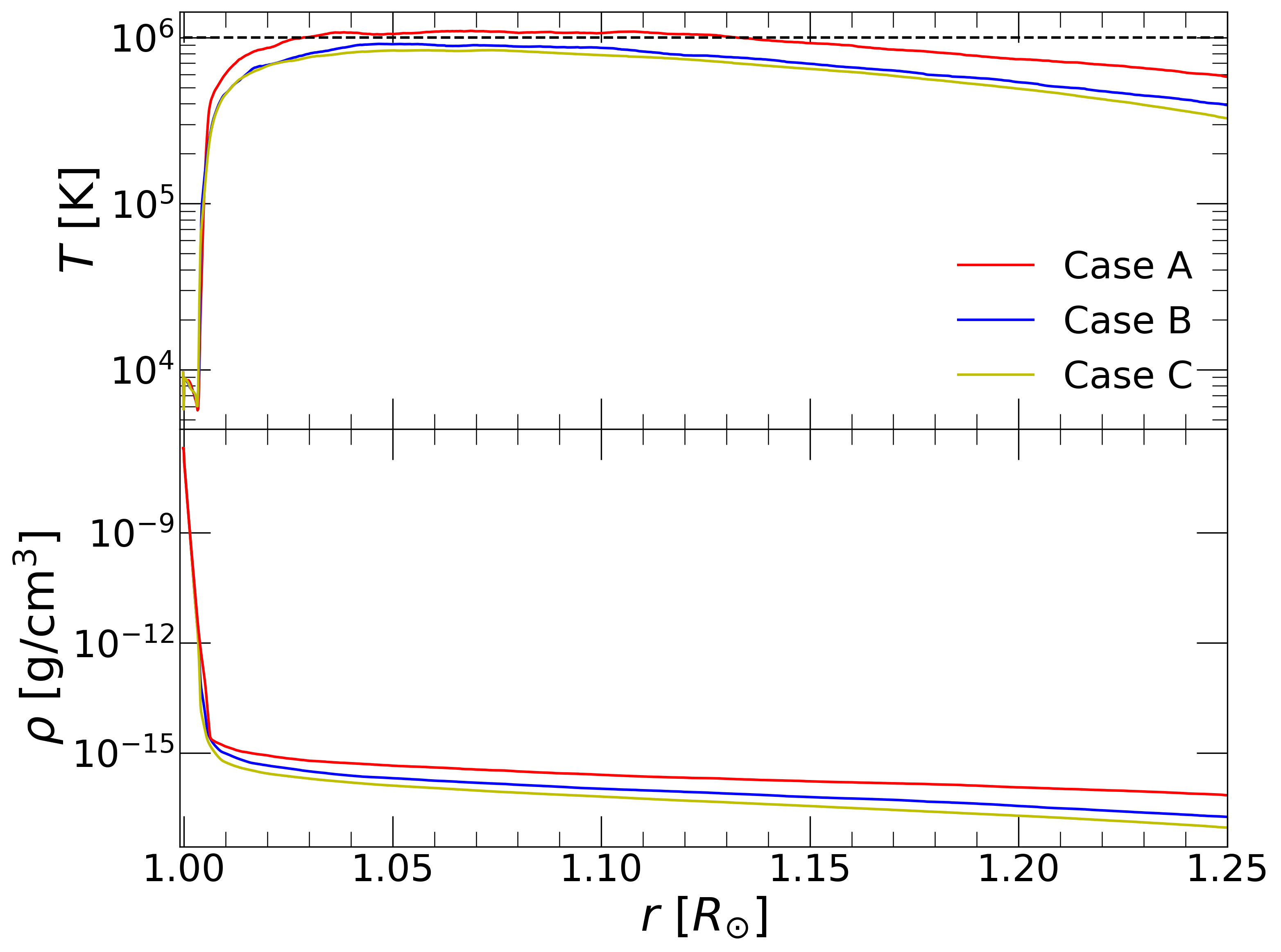}
	\caption{Mean density and temperature profiles, averaged over the last 100 days (from Day 265 to Day 365) for the three cases.}
	\label{fig:meanTrho}
\end{figure}

Observationally, the mean temperature in the quiet solar corona was 1.4--1.8 MK~\cite{Morgan2017}. While Figure \ref{fig:last_50_days} indicates that the coronal temperature in the simulation could occasionally reach this range, this does not guarantee that the mean temperature can be maintained in this range. To test this, we calculated the mean temperature and density profiles of the simulations, averaged over the last 100 days for the three cases, as shown in Figure \ref{fig:meanTrho}. We can see that as $k_0$ increased, both the mean temperature and the mean density in the simulated corona increased. As mentioned above, this is to be expected, because when $k_0$ increases, the heating flux in the chromosphere increases and consequently more material with higher temperature is driven into the corona.
For the optimal simulation, Case B, the maximum mean temperature was only 0.92 MK (see Table \ref{tab:3cases}), much lower than the observed mean temperature in the quiet corona. Even with a $k_0$ ten times larger (Case A), the mean temperature could only reach 1.1 MK, still lower than the observed value. We can also clearly see in Figure \ref{fig:meanTrho} that the mean temperature dropped monotonically as the height increased past the location corresponding to the maximum value, due to the quick dissipation of acoustic shock waves. Both phenomena indicate that heating via acoustic shock waves alone cannot maintain a solar corona as observed, and other heating mechanisms, e.g., magnetic reconnection and the dissipation of Alfv{\'e}nic waves, must play an important role in coronal heating.

%=========================================
\subsection{Origin of Global 4 {mHz} Oscillation}\label{shockwave}

In our model, the chromosphere is sustained by the artificial chromospheric energy flux, which is also the energy source for maintaining the corona, as our model will collapse when $k_0$ is less than its critical value, instead of exhibiting a corona without the chromosphere. This also implies that the coronal heating in our model must require the existence of the chromosphere. We can approximately locate the birthplace of the coronal heating shock waves in Figures \ref{fig:rho_T_Vr_profiles} and \ref{fig:last_50_days} between the top of the chromosphere and the bottom of the corona. In this section, we investigate these shock waves in further detail. Below, we focus on a discussion of Case B, while the relevant results for the other two cases are shown in Table \ref{tab:3cases}.

\begin{figure}[H]
%	\centering
	\includegraphics[width=1.0\linewidth]{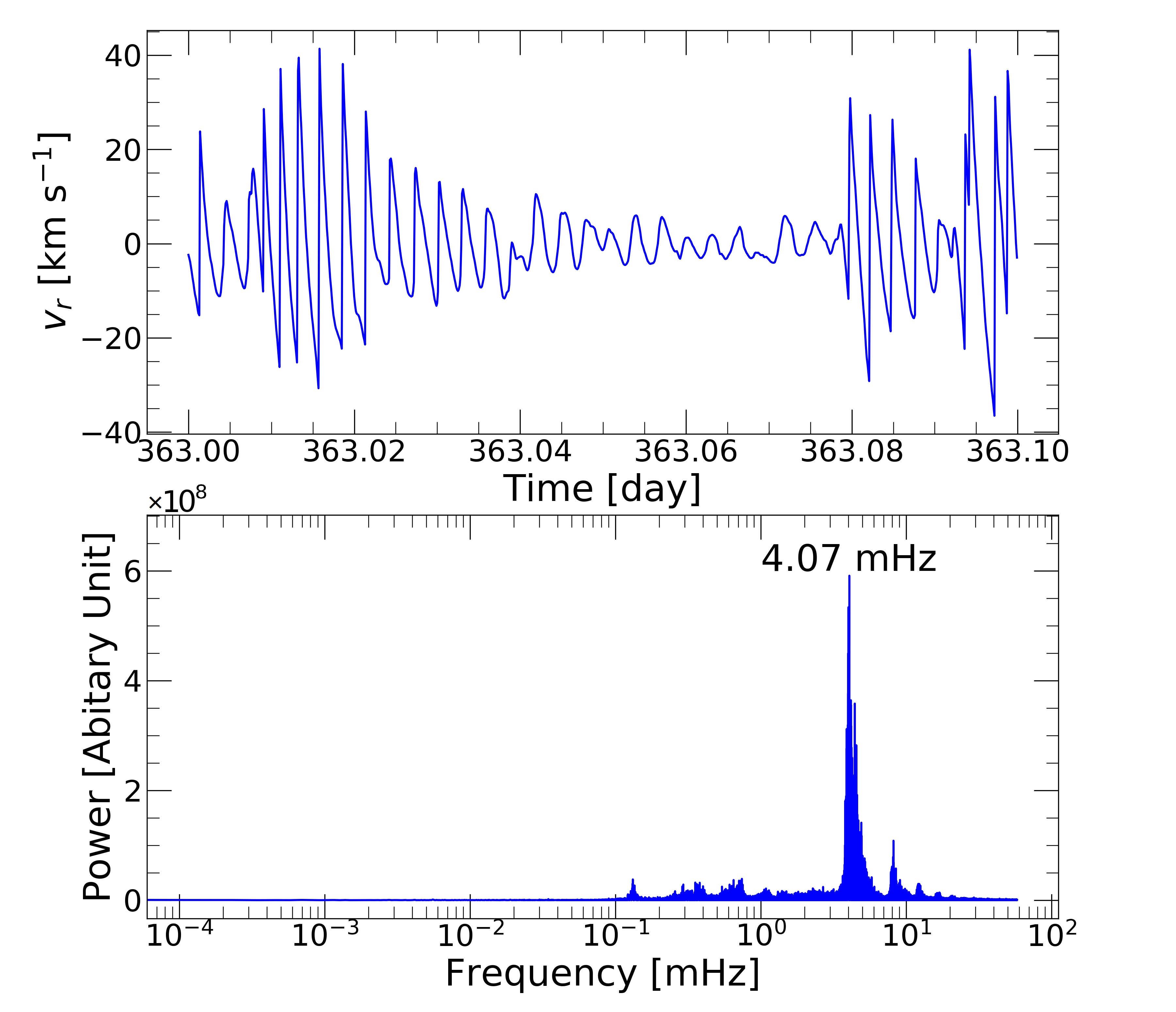}
	\caption{Profile %MDPI: Please change the terms into scientific notations in the figure,  e.g., "$8 \times 10^{3}$", not "8e3".
 of $v_r$ over time sampled at a fixed height of $1.5\times10^4$ km above the photosphere (upper panel) and its PSD (lower panel), for Case B. Only data from Day 363 to 363.1 are displayed in the upper panel, to show more details. The PSD in the lower panel was calculated with data from Day 265 to 365, sampled  with a time-step of $10^{-4}$ day. The peak frequency in the lower panel was $4.07$ mHz, denoted by a vertical dashed line.}
	\label{fig:freq_of_shock_train}
\end{figure}

\begin{figure}[H]
%	\centering
	\includegraphics[width=1.0\linewidth]{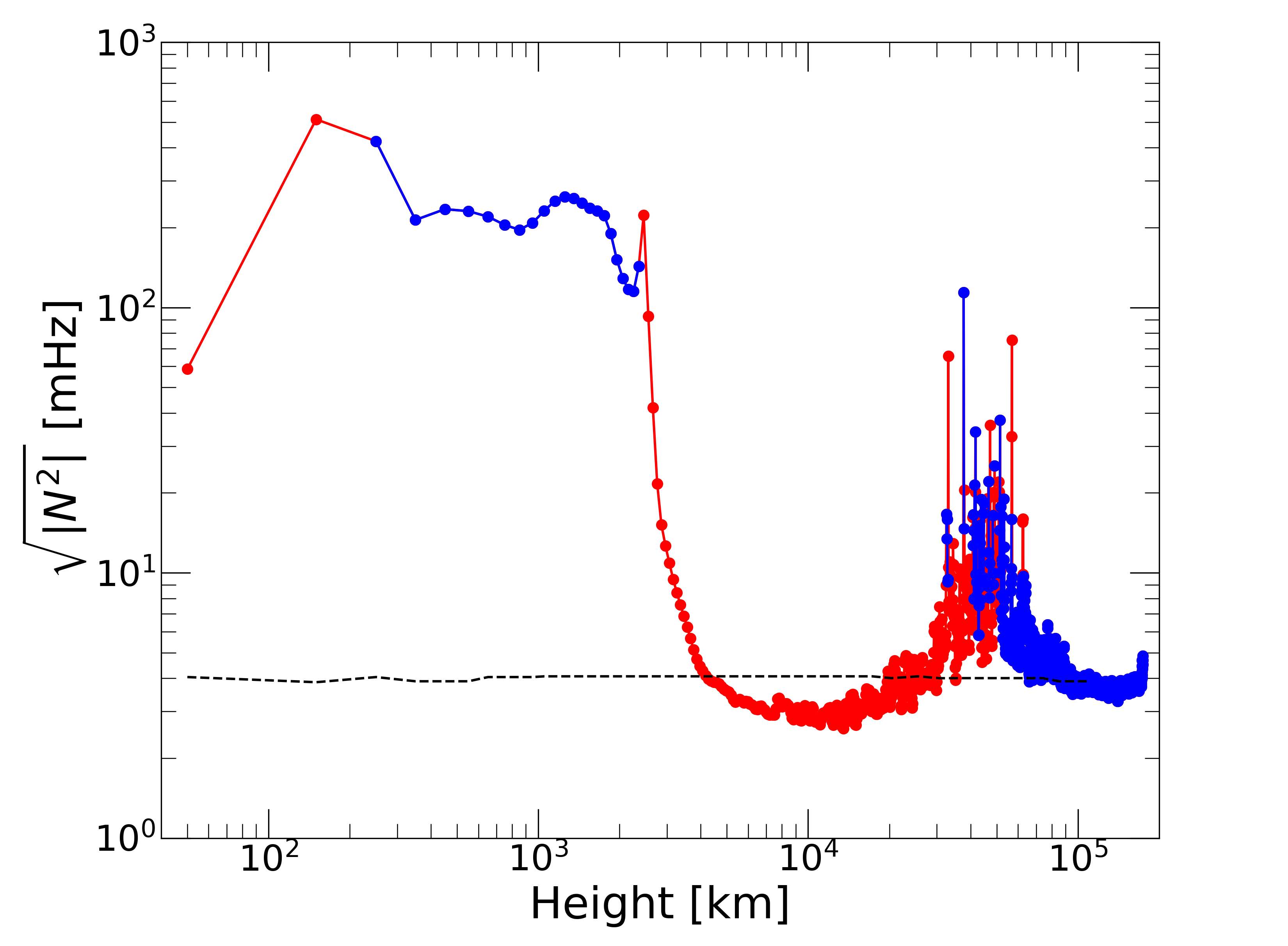}
	\caption{The profile of critical frequencies at different heights for Case B. Red dots were calculated from positive $N^2$ and blue dots from negative $N^2$. Acoustic waves with frequencies higher than the red dots can freely propagate at corresponding heights, while those lower than the red dots are rapidly damped. The heights with blues dots always have large damping. The dashed line represents the peak frequency in the PSD of the $v_r$ profile at each corresponding height.}
	\label{fig:curve_of_critfreq}
\end{figure}

First, we measured the frequency of successive shock waves in our model and compared it with observations. We sampled the instantaneous radial velocities $v_r$ at different heights over the last 100 days of our simulation (Day 265 to Day 365) with a time-step of $10^{-4}$ day. The data at each height formed a fluctuating velocity profile, as presented in the upper panel of Figure \ref{fig:freq_of_shock_train}, which is the $v_r$ profile at the height of $1.5 \times 10^4$ km above the photosphere, and only the data from Day 363 to 363.1 are displayed to show more details. The PSD of the $v_r$ profile was then calculated with fast Fourier transformation at each height, as presented in the lower panel of Figure \ref{fig:freq_of_shock_train}. The peak frequency in the PSD, which can either arise from the local $p$-mode oscillations or correspond to the periodic fluctuation due to the passing of acoustic waves or successive shock waves, was 4.07 mHz at the height of $1.5 \times 10^4$ km (see Figure \ref{fig:freq_of_shock_train}), and lay in the range of 3.9--4.07 mHz at different heights (denoted by the dashed line in Figure \ref{fig:curve_of_critfreq}), consistent with the observed typical wave periods in the corona of around three to five minutes ($3.33$--$5.55$ mHz;~\cite{DeMoortel2002, VanDoorsselaere2008, Tomczyk2009, Morton2016, Morton2019}). It should be noted that, as the height increased, while the peak at $\sim$4 mHz always existed in the PSD except for places close to the outer boundary at $r=1.25R_\odot$ (height of $\sim$1.74 $\times 10^5$ km), another peak at $\sim$0.3 mHz appeared and gradually became stronger, which we suspect also resulted from boundary effects.

Second, we investigated the propagation of acoustic waves in our model atmosphere.
According to the linearized perturbation theory of stellar oscillations (e.g.,~\cite{Cox1980, Unno1989}), only $p$-mode oscillations exist in our spherically symmetric model. The $p$-mode oscillations excite acoustic waves with the same frequency, which can be investigated through the Brunt-V\"{a}is\"{a}l\"{a} critical frequency, above which waves can freely propagate, rather than rapidly damping in the surroundings, which is calculated as~\cite{Unno1989}
\begin{equation}
N^2=\left(\frac{GM_{\odot}}{r^2}\right)^2\frac{\rho}{p}\left[\nabla_{\mu}+\frac{4-3\beta}{\beta}\left(\nabla_{\rm{ad}}-\nabla\right)\right],
\end{equation}
where $N$ is the critical frequency, $\beta$ is the ratio of gas to total pressure, $\nabla_{\mu}$ represents the inhomogeneity of elements, $\nabla$ is defined as $\nabla \equiv {d\ln T}/{d\ln p}$, and $\nabla_{\rm{ad}}$ is the value of $\nabla$ in the adiabatic case. 
Note that $N^2$ can be either positive or negative here, where positive values signify that the propagation of acoustic waves with frequencies no less than $\sqrt{\left|N^2\right|}$ is permitted in the relevant computing cells, and negative values correspond to places where the waves are evanescent and non-propagating. In our calculation, we ignored the inhomogeneity of elements ($\nabla_{\mu}=0$) and radiation pressure ($\beta=1$), and assumed $\nabla_{\rm{ad}}=0.4$, which is the adiabatic value of a fully ionized ideal gas. Then, we calculated $N^2$ with the mean values of density, pressure, and temperature from Day 265 to Day 365, as presented in Figure \ref{fig:curve_of_critfreq}. The dots represent the critical frequencies $\sqrt{\left|N^2\right|}$ calculated at the center of the computing cells. Red dots were calculated from positive $N^2$, and blue dots from negative $N^2$. The dashed line represents the peak frequency in the PSD of the $v_r$ profile at each corresponding height (see Figure \ref{fig:freq_of_shock_train} as an example). We can see that the region below $3\times10^3$ km, including the whole chromosphere, was a cut-off layer for acoustic waves, where $N^2$ were mostly negative and even the places with positive $N^2$ only allowed the propagation of acoustic waves with frequencies much larger than the peak frequencies sampled from $v_r$ profiles. Thus, it is impossible that the $\sim$4 mHz coronal heating shock waves identified in our model came from the photosphere (in fact, our model did not inject any waves from the photosphere boundary). The region above $4\times10^4$ km also had large damping for the propagation of acoustic waves, with mostly negative $N^2$, though the critical frequencies in the limited places with positive $N^2$ in this region were lower than those in the model chromosphere. On the other hand, the region between $\sim$4$\times10^3$ and $\sim$2$\times10^4$ km, which lies at the bottom of the corona, allowed the free propagation of acoustic waves of $\sim$4 mHz. We conjecture that this region is a resonant cavity for the acoustic waves excited by $p$-mode oscillations, which amplifies the oscillations and gives birth to shock waves. Although the acoustic waves cannot propagate stably far outside of this region, shock waves of the same frequency form in this region and successively carry energy to the higher corona, which can be seen in Figure \ref{fig:rho_T_Vr_profiles}.

Alfv{\'e}nic waves have been ubiquitously observed (both in space and time) in the solar corona~\cite{Tomczyk2007, McIntosh2011, Morton2016, Morton2019}. Morton et al. (2019)~\cite{Morton2019} found an enhancement in power around 4 mHz (mean and standard error of mean are $4.0 \pm 0.1$ mHz, with a standard deviation of $1$ mHz) in the velocity PSD of Alfv{\'e}nic waves, both from the Coronal Multi-channel Polarimeter (CoMP) and the {Solar Dynamics Observatory (SDO)} Atmospheric Imaging Assembly (AIA) data. This enhancement exists in a large majority (>95 per cent) of coronal power spectra (i.e., throughout the entire corona) and is almost invariant in the center values throughout the solar cycle. Traditionally, it was proposed that Alfv{\'e}nic waves are generated in the solar photosphere (e.g.,~\cite{Cranmer2005}), though the low-ionization fraction in the photosphere hinders the generation~\cite{Vranjes2008} (see~\cite{Tsap2011} for a different view), and the transmission of Alfv{\'e}nic waves also suffers from a strong reflection at the {TR}. Several {MHD} wave models have been proposed to solve this problem (e.g.,~\cite{Cally2008,Cally2011,Felipe2012,HC2012,Cally2017}), where Alfv{\'e}nic waves can be excited by helioseismic $p$-mode oscillations through double mode conversion and the TR reflection can be greatly reduced under certain conditions. However, these models need to be fine-tuned to provide sufficient Alfv{\'e}nic waves as observed in the corona, and may have difficulty in explaining the ubiquitous existence of the Alfv{\'e}nic waves peaked $\sim$4 mHz. On the other hand, our hydrodynamic simulation found that acoustic waves were newly excited in a resonant cavity at the bottom of the corona, so that when magnetic fields are taken into consideration, Alfv{\'e}nic waves could be generated there with processes similar to the excitation of Alfv{\'e}nic waves in the photosphere or through double mode conversion, while naturally overcoming the traditional problems of the low-ionization fraction in the photosphere and the strong TR reflection. The acoustic waves in our model also peaked at $\sim$4 mHz in the PSD, which coincides with the peak in the observed PSD of Alfv{\'e}nic waves. {The energy flux of acoustic waves in our model can be estimated as $\sim$2.84 $\times 10^5$~$\rm{erg\cdot cm^{-2}\cdot s^{-1}}$, which is similar to the energy flux of observed Alfv{\'e}nic waves of $\sim$1.54 $\times 10^5$~$\rm{erg\cdot cm^{-2}\cdot s^{-1}}$ \endnote{{We used Equation (\ref{eq_flux}) to calculate the estimation of the acoustic energy flux. The quantities required in the equation were obtained from the profiles of Day 354 (blue lines in Figure \ref{fig:rho_T_Vr_profiles}) at $r=0.004R_{\odot}$ ($\sim$2800 {km}, the birthplace of acoustic waves). They are the density $\rho\simeq10^{-14}\rm{g}$, temperature $T\simeq10^5\rm{K}$, and velocity amplitude $v_a\simeq25$~$\rm{km\cdot s^{-1}}$, respectively. The energy flux of Alfv{\'e}nic waves was obtained from Supplementary Table 2 of~\cite{Morton2019}.}}. These imply that the acoustic waves generated in our model may be sufficient to become the driving source of the Alfv{\'e}nic waves observed by Morton et al. (2019)~\cite{Morton2019}.}

De Pontieu et al. (2004)~\cite{Pontieu2004} proposed that solar chromospheric spicules are driven by the leakage of photospheric $p$-mode oscillations, which can be achieved through the waveguide of magnetic flux tubes. Their model can well explain the observed spicule periodicity, $350\pm60$ s (frequency of $2.86\pm0.59$ mHz). This frequency is similar to that of the shock train in our model ($\sim$4 mHz). Their simulations showed that the oscillations leaking into the chromosphere can develop non-linearly into upward propagating shocks with a velocity amplitude $\sim$20--40 {km/s}, which is also similar to our model (see the panels showing velocity profiles in Figures \ref{fig:rho_T_Vr_profiles} and \ref{fig:last_50_days}). Thus, we think that our model can be regarded as an extension of their model, especially regarding their omission of coronal heating. 

%=========================================
\subsection{Mass Upflows}\label{upflows}

\begin{figure}[H]
%	\centering
	\includegraphics[width=1.0\linewidth]{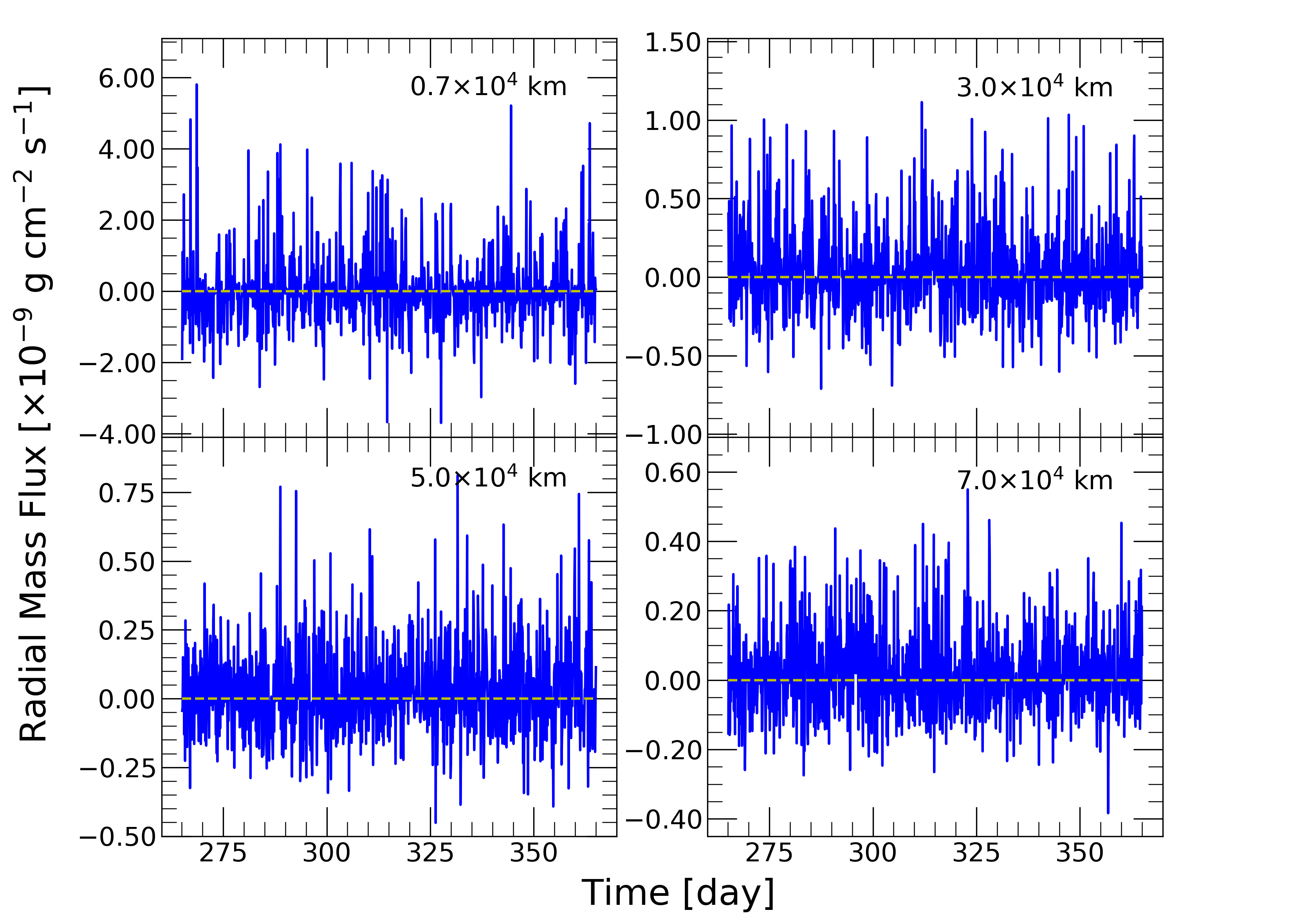}
	\caption{Radial mass fluxes at four different heights (solid lines) and their {zero-flux levels} (dashed lines) for Case B, recorded from Day 265 to Day 365.}
	\label{fig:mass_flux}
\end{figure}

Figure \ref{fig:mass_flux} displays the radial mass fluxes of Case B at four different heights (solid lines) from Day 265 to Day 365, and {the dashed line in each panel denotes the zero-flux level as the reference line of radial outflow (positive) and inflow (negative).} While the oscillations of mass fluxes are natural results of shocks waves, we can see that the absolute values of mass fluxes at the peaks are much larger than those at the troughs, indicating that mass upflows are superposed onto the peaks. These mass upflows occurred in our model in the height range $\sim$7 $\times 10^3$--$7\times10^4$ km, which coincided with the observed height of solar prominences ($\sim$$10^4$--$10^5$ km,~\cite{Aschwanden2005}). While prominences are believed to be a kind of magnetic field structure formed in filament channels, how their mass is acquired is still an open question. Aschwanden (2005)~\cite{Aschwanden2005} (also see~\cite{Parenti2014}) proposed three possible scenarios, one of which is injection by chromospheric mass upflows, and there is a lot of observational evidence that mass is continuously entering and exiting the filament magnetic field throughout its lifetime. These features agree quite well with the mass upflows in our model, so we conjecture that they might be the mass supplier of prominences, which could later evolve in magnetic fields into the observed prominence structure, though these structures could not be obtained in our simulation as we neglected magnetic fields in the corona. These mass upflows also existed for the other two cases, though the exact height range differed slightly.

%%%%%%%%%%%%%%%%%%%%%%%%%%%%%%%%%%%%%%%%%%
\section{Summary and Conclusions}\label{Sec:conclusions}

We performed global spherically symmetric HD simulations of the solar atmosphere, to explore the origin of global $4\ \rm{mHz}$ oscillations in corona and investigate the possibility of corona heating via the resulting shock waves of these oscillations. The range of the simulation was set between 1 and 1.25 $R_{\odot}$,  so that the chromosphere and corona could evolve as a whole. The heating in the chromosphere was simplified by an artificial energy flux proportional to local pressure, $F_{\rm c} = k_0 p$ (applied in the height range 0--2000 km above the photosphere; see Section \ref{Sec:ch_flux} for its physical interpretation), which simulated the heating flux of acoustic waves in the chromosphere without adding an acoustic wave generator in the photosphere boundary. We used this artificial flux for two reasons. First, with $F_{\rm c}$ we could ensure that the acoustic waves in the simulated corona were newly generated, instead of being leaked from the photosphere boundary. Secondly, the simulated chromosphere could quickly converge to a quasi-steady state with this form of $F_{\rm c}$. While this assumption is very simple, the artificial energy fluxes in the simulation actually agreed quite well with the observed acoustic energy fluxes (Figure \ref{fig:artificial_flux}) for the optimal value of $k_0= 1.61\times10^3\rm{cm\ s^{-1}}$ (Case B) that we found. The mass loss rate at the outer boundary also agreed with observed value, and we obtained a chromospheric structure which roughly agreed with the VAL-C model determined from \textit{Skylab} observations of the quiet Sun (see Figure \ref{fig:num_vs_obs}). Therefore, we think that our simple assumption reflects part of the physical nature of the solar chromosphere, and that the simulation results are valid. {However, it should be noted that we ignored the heat conduction in our model, which is important, especially within the TR. According to our computational practice, the effect of ignoring heat conduction did not alter the generation of $4\ \rm{mHz}$ oscillations appearing in the early stages of our simulations, but we cannot remark about its long-term effects through comparing runs with and without the heat conduction.}

We found that shock waves are produced from acoustic waves excited by $p$-mode oscillations at the bottom of the corona and propagate upwards, heating the coronal plasma along the way, occasionally up to >1 MK. However, the maximum mean temperature was only 0.92 MK for  the optimal simulation, Case B. Even with a $k_0$ ten times larger (Case A; the larger $k_0$, the larger the heating flux in the chromosphere), the mean temperature could only reach 1.1 MK. As a comparison, the observed mean temperature in the quiet solar corona is 1.4--1.8 MK~\cite{Morgan2017}. The mean temperature also dropped monotonically as the height increased past the location corresponding to the maximum value, due to the quick dissipation of acoustic shock waves. Both phenomena indicate that the heating via acoustic shock waves alone cannot maintain a solar corona as observed, and other heating mechanisms such as magnetic reconnection and the dissipation of Alfv{\'e}nic waves must play an important role in coronal heating. However, it should be noted that this kind of shock wave heating may still have been exaggerated, due to the one-dimensional nature of our simulations.

For Case B, the frequencies of the shock waves were found to be in the range of 3.9--4.07 mHz (see Table \ref{tab:3cases} for the results of the other two cases), in agreement with typical wave periods observed in the corona, which peak at around three to five minutes~\cite{DeMoortel2002, VanDoorsselaere2008, Tomczyk2009, Morton2016, Morton2019}. We also investigated the Brunt-V\"{a}is\"{a}l\"{a} cut-off frequency for acoustic waves at different heights. While the regions below the height of $3\times10^3$ km and above the height of $4\times10^4$ km were largely opaque for acoustic waves of $\sim$4 mHz,  the region between $\sim$$4\times10^3$ and $\sim$$2\times10^4$ km allowed their free propagation. Thus, this region could act as a resonant cavity for $p$-mode oscillations of $\sim$4 mHz and give birth to shock waves of the same frequency, which could
then carry energy into the higher corona {with a speed $\sim$$139\ \rm{km\cdot s^{-1}}$}. As the source of acoustic waves involved in the coronal heating is moved from the photosphere to the top of the chromosphere, our model naturally overcomes the problem of chromospheric damping and blocking of acoustic waves. The amplitude of the coronal heating shocks in our model is in the order of tens of kilometers per second in the $v_r$ profile, much larger than that required by chromospheric heating (a few kilometers per second, see~\cite{Zhang2021}), while the density in the corona is much smaller than that in the chromosphere, so it is natural that the model corona can be heated to a much higher temperature. Note that our simulation was purely hydrodynamic and did not include the effects of magnetic fields. When magnetic fields are added, as in the real solar atmosphere, the energy stored in acoustic waves can be converted into other forms, such as Alfv{\'e}nic waves, spicules, and prominences, which explains why such high amplitude acoustic waves are not typically observed.

Our model also helps to explain the ubiquitously (both in space and time) observed Alfv{\'e}nic waves in the solar corona. Morton et al. (2019)~\cite{Morton2019} found an enhancement in power around 4 mHz in the PSD of Alfv{\'e}nic waves throughout the entire corona, which was almost invariant in the center values throughout the solar cycle. Traditionally, it was proposed that Alfv{\'e}nic waves are generated in the solar photosphere, though the low-ionization fraction in the photosphere hinders the generation, and the transmission of Alfv{\'e}nic waves suffers from a strong TR reflection. Several MHD wave models have been proposed to solve this problem (e.g.,~\cite{Cally2008,Cally2011,Felipe2012,HC2012,Cally2017}), where the Alfv{\'e}nic waves can be excited by helioseismic $p$-mode oscillations through double mode conversion and the TR reflection can be greatly reduced under specific conditions. However, these models need to be fine-tuned to provide sufficient Alfv{\'e}nic waves as observed in the corona, and may have difficulty in explaining the ubiquitous existence of the Alfv{\'e}nic waves peaking at $\sim$4 mHz. On the other hand, our hydrodynamic simulation found that acoustic waves are newly excited in a resonant cavity at the bottom of the corona, so that when magnetic fields are taken into consideration, Alfv{\'e}nic waves can be generated there with processes similar to the excitation of Alfv{\'e}nic waves in the photosphere or through double mode conversion, while naturally overcoming the traditional problems of the low-ionization fraction in the photosphere and the strong TR reflection. The acoustic waves in our optimal simulation (Case B) also peaked at $\sim$4 mHz in the PSD, which coincides with the peak in the observed PSD of Alfv{\'e}nic waves.

We also found that mass upflows existed in the height range $\sim$$7\times10^3$--$7\times10^4$ km for Case B, which might be the mass supplier of solar prominences. These mass upflows also existed for the other two cases, though the exact height range differed slightly.

Finally, it is necessary to note the theoretical rationality of our results in this paper. We think that our scenario of general 4 mHz oscillations in solar corona is physically reasonable, but its accompanying acoustic shock heating may still be exaggerated due to the one-dimensional nature of our model. Shock waves can only propagate radially outward in our model, which allows them to be enhanced through superposition during collisions in the same direction, resulting in greater model coronal heating. On the contrary, real shock waves can travel in any directions in the real solar atmosphere, which makes the superposition enhancement less efficient~\cite{Ulmschneider_2005}. However, we still think that this kind of heating may have significant effects in certain special one-dimensional situations, such as in magnetic flux tubes, but not in global solar corona. 

%%%%%%%%%%%%%%%%%%%%%%%%%%%%%%%%%%%%%%%%%%
\vspace{6pt} 

%%%%%%%%%%%%%%%%%%%%%%%%%%%%%%%%%%%%%%%%%%
%% optional
%\supplementary{The following supporting information can be downloaded at:  \linksupplementary{s1}, Figure S1: title; Table S1: title; Video S1: title.}


%\begin{tabularx}{\textwidth}{lll}
%\toprule
%\textbf{Name} & \textbf{Type} & \textbf{Description} \\
%\midrule
%S1 & Python script (.py) & Script of python source code used in XX \\
%S2 & Text (.txt) & Script of modelling code used to make Figure X \\
%S3 & Text (.txt) & Raw data from experiment X \\
%S4 & Video (.mp4) & Video demonstrating the hardware in use \\
%\bottomrule
%\end{tabularx}
%}

\vspace{6pt}  
%%%%%%%%%%%%%%%%%%%%%%%%%%%%%%%%%%%%%%%%%%
\authorcontributions{Conceptualization, L.X. and C.J.; methodology, L.X. and C.J.; software, L.X. and L.Z.;  writing---original draft preparation, L.X. and C.J.; writing---review and editing, C.J. and L.X.; visualization, L.X. and L.Z.; funding acquisition, L.X. and C.J. All authors have read and agreed to the published version of the manuscript.}

\funding{This %MDPI: Information regarding the funder and the funding number should be provided. Please check the accuracy of funding data and any other information carefully.
 work is supported by the National Key R\&D Program of China under grant No. 2023YFA1607902, the National Natural Science Foundation of China under grants No. 11703083 and 12221003, and the Natural Science Foundation of Fujian Province of China under grant No. 2023J01008.}

\dataavailability{The data and codes underlying this article will be shared on reasonable request to the corresponding authors.} 

% Only for journal Nursing Reports
%\publicinvolvement{Please describe how the public (patients, consumers, carers) were involved in the research. Consider reporting against the GRIPP2 (Guidance for Reporting Involvement of Patients and the Public) checklist. If the public were not involved in any aspect of the research add: ``No public involvement in any aspect of this research''.}

% Only for journal Nursing Reports
%\guidelinesstandards{Please add a statement indicating which reporting guideline was used when drafting the report. For example, ``This manuscript was drafted against the XXX (the full name of reporting guidelines and citation) for XXX (type of research) research''. A complete list of reporting guidelines can be accessed via the equator network: \url{https://www.equator-network.org/}.}

% Only for journal Nursing Reports
%\useofartificialintelligence{Please describe in detail any and all uses of artificial intelligence (AI) or AI-assisted tools used in the preparation of the manuscript. This may include, but is not limited to, language translation, language editing and grammar, or generating text. Alternatively, please state that “AI or AI-assisted tools were not used in drafting any aspect of this manuscript”.}

\acknowledgments{We thank Tong %MDPI: Titles (e.g., Dr., Mr., and Prof.) should NOT be used in the Acknowledgments section. We removed them. Please confirm.
 Liu for helpful discussion.}

\conflictsofinterest{The authors declare no conflicts of interest. The funders had no role in the design of the study; in the collection, analyses, or interpretation of data; in the writing of the manuscript; or in the decision to publish the results.} 

%%%%%%%%%%%%%%%%%%%%%%%%%%%%%%%%%%%%%%%%%%
%%%%%%%%%%%%%%%%%%%%%%%%%%%%%%%%%%%%%%%%%%
\begin{adjustwidth}{-\extralength}{0cm}
\printendnotes[custom] % Un-comment to print a list of endnotes

\reftitle{References}

\PublishersNote{}
\end{adjustwidth}
\end{document}